\documentclass[11pt,doublespace,fullpage]{article}

\usepackage{amsmath}
\usepackage{comment,bm}
\usepackage{amssymb,latexsym,graphicx}
\usepackage{natbib}
\usepackage{setspace}
\usepackage{adjustbox}
\usepackage{xcolor,multirow}
\usepackage{hyperref}
\hypersetup{allcolors=black, allbordercolors=white}
\usepackage{mathptmx}

\usepackage{subfig}
\usepackage{graphicx}
\usepackage{xcolor}
\usepackage{boxedminipage}
\usepackage{amsfonts,amsmath,amssymb}
\usepackage{comment,bm}
\usepackage{helvet, multirow, multicol}
\usepackage{titlesec}
\usepackage{longtable}
\usepackage[figuresright]{rotating}
\usepackage{adjustbox,url,natbib}

\def\E{\mathbb{E}}

\def\by{\bm{y}}
\def\bY{\bm{Y}}

\def\bI{\bm{I}}

\def\E{\mathbb{E}}

\def\E{\mathbb{E}}

\def\MN{\mathbf{N}}

\def\bo{\bm{1}}

\begin{document}

\title{\bf Efficient and powerful equivalency test on combined
mean and variance with application to diagnostic device comparison studies}
\author{Yun Bai$^{1}$, Zengri Wang$^{1}$, Theodore Lystig$^{1}$, Baolin Wu$^{2}$
\thanks{Email: baolin@umn.edu} \\ 
$^1$Medtronic plc, Minneapolis, MN, USA \\
$^2$Division of Biostatistics, School of Public Health, University of Minnesota}
\date{}
\maketitle

\begin{abstract}
In medical device comparison studies, equivalency test is commonly
used to demonstrate two measurement methods agree up to a pre-specified performance goal
based on the paired repeated measures. Such equivalency test often
involves controlling the absolute differences that depend on both the mean and variance parameters,
and poses some challenges for statistical analysis.
For example, for the oximetry comparison study that motivates our research, FDA has clear guidelines approving an investigational 
pulse oximeter in comparison to a standard oximeter via testing the root mean squares (RMS), a composite measure of both mean and variance parameters.  
For the hypothesis testing of this composite measure, existing methods have been either exploratory or relying on the 
large-sample normal approximation with conservative and unsatisfactory performance.
We develop a novel generalized pivotal test to rigorously and accurately test 
the system equivalency based on RMS. The proposed method has well-controlled type I error and favorable 
performance in our extensive numerical studies.
When analyzing data from an oximetry comparison study, 
aiming to demonstrate performance equivalency between an FDA-cleared oximetry system and an investigational system, our proposed method resulted in a highly significant test result strongly supporting the system equivalency. 
We also provide efficient R programs for the proposed method in a publicly available R package.
Considering that many practical equivalency studies of diagnostic devices are of small to medium sizes,
our proposed method and software timely bridge an existing gap in the field. 
\end{abstract}

\section{Introduction}

In medical device comparison studies, equivalency test is commonly
used to demonstrate two measurement methods agree up to a pre-specified performance goal
based on the paired repeated measures. 
For example, for the oximetry comparison study that motivates our research, 
FDA has clear guidelines on how to establish the 
acceptance of a new investigational oximeter in comparison to a standard oximeter 
through data collection and a formal hypothesis testing framework
based on testing the root mean squares (RMS), which is a composite measure of both mean and variance parameters.  
However existing methods are mostly exploratory \citep{lin_concordance_1989,lin_assay_1992,bland1999,bland2007}
or have relied on the large-sample normal approximation with conservative and unsatisfactory performance
\citep{Pennello2002,Pennello2003,ndikintum2016}.

In this paper, we develop statistical methods for the equivalency test problem in the diagnostic device comparisons.
We offer novel insights to the statistical problems with respect to the equivalency test
and provide accurate and efficient solutions for practical use. 
Specifically 
we develop powerful statistical methods with robust statistical properties. 
  The newly developed methods are motivated by the novel generalized pivotal statistics based approach \citep{weerahandi1995}. 
  Through extensive numerical studies, we demonstrate favorable performance of the proposed methods:
  they achieve higher power than the existing methods by a large margin, and can produce accurate confidence 
  intervals to precisely quantify the variation of key parameter of interest.
We further develop a publicly available R package that implements the new methods. 
Our proposed statistical methods along with the R package provide
useful practical tools to the general community and timely bridge the gap in the field.

The rest of the paper proceeds as follows. Section 2 explains the root mean squares (RMS) testing problem in a 
linear random-effects model and our proposed method based on a generalized inference approach. 
Next, we study the performance of our method through simulation studies in Section 3 and
apply the new method to the pulse oximetry data in Section 4.  A discussion is given in Section 5.
Throughout the paper, we will mainly focus on intuitive ideas and delegate all technical details to the
Appendix.

\section{Statistical methods}

Consider a random-effects ANOVA model, $Y_{ij}=\mu+u_i+\epsilon_{ij}$, $i=1,\cdots,n;j=1,\cdots,m_i$,
where $\mu$ is the overall mean, the random effects term $u_i\sim\MN(0,\sigma_b^2)$ and the
random error components $\epsilon_{ij}\sim\MN(0,\sigma_w^2)$ are all independent. 
For paired repeated measure data, $Y_{ij}$ will describe the differences of two measures
(e.g. pulse oximeter measurement and the co-oximeter measurement).
Denote $\bY_i=(Y_{i1},\cdots,Y_{im_i})$,
and $\bY=(\bY_1,\cdots,\bY_n)$. We can readily check that 
$Var(Y_{ij}) = \sigma_b^2 +\sigma_w^2$, $Cov(Y_{ij},Y_{ik}) = \sigma_b^2$.
Here $\sigma_b^2$ is the between-level variance, $\sigma_w^2$ is the within-level variance, and $\mu$ quantifies the average mean value. This is a special case of the linear mixed effects model (LMM; \citealp{lmm1982}).

In the following, we use capital letters to denote the random variables, e.g. $(Y_{ij},\bY_i,\bY)$,
and the corresponding small letters to denote the observed data, e.g. $(y_{ij},\by_i,\by)$.

Our main interest is the root mean squares (RMS) parameter defined as $\rho=\sqrt{\mu^2+\sigma_b^2+\sigma_w^2}$.
In the paired repeated measure comparison, $\rho$ quantifies the average
absolute difference between the two measures, since  $\E(Y_{ij}^2)=\rho^2$.
It is of interest to test $H_0: \rho \geq \rho_0$ versus
$H_a: \rho <\rho_0$, where $\rho_0$ is a pre-specified acceptable threshold.
In addition, we want to compute a confidence interval (CI) for $\rho$ to quantify its variation.

Per the FDA guideline, a pulse oximeter is approvable if the null hypothesis
$H_0: \rho\geq \rho_0$ can be rejected in favor of the alternative hypothesis
$H_a: \rho<\rho_0$ with a pre-specified value $\rho_0$. For example $\rho_0=3\%$
is often used for transmittance, wrap and clip pulse sensors.

The RMS parameter $\rho$ involves both the mean parameter $\mu$ and variance parameters
$(\sigma_b^2,\sigma_w^2)$. 
It poses significant challenges to construct exact methods for inference.
To our knowledge, this has not been well-studied in the filed. 
Partly owing to the composite nature of $\rho$ and the presence of nuisance parameters,
the large-sample based normal approximation 
generally does not perform well \citep{Pennello2002,Pennello2003,ndikintum2016}.
We note that the statistical inference of individual mean and variance parameters
has been well-studied, and a variety of methods exist
\citep{MLS1990,zhou1994,park2003,MLS2004,S2CI2004,MuCI2004,burdick2006}.  
In the following we adopt the method of generalized inference \citep{GP1989,GCI1993} 
to conduct significance test and calculate CI for $\rho$.

\subsection{Generalized test statistic} 

In the presence of nuisance parameters,
the key idea of generalized inference is to construct a generalized pivotal quantity/test statistic, 
which is defined as a function of observed data, parameters and random variables.
It has a known distribution free of any unknown parameters, and hence can be used to conduct
hypothesis test and calculate CI for the parameter of interest. 
Compared to the traditional statistical inference based on the pivotal statistic, 
which is defined as the function of parameters and random variables with a known distribution, 
the generalized inference allows the incorporation of observed data.
We refer the reader to the monograph of \cite{weerahandi1995}
for a detailed discussion along with relevant examples. 

Denote the summary statistics, \
$\bar{Y}_i = \sum_{j=1}^{m_i}Y_{ij}/m_i$, and
$S_i^2 = \sum_{j=1}^{m_i}(Y_{ij}-\bar{Y}_i)^2/(m_i-1)$.
We note that $\bar{Y}_i\sim\MN(\mu,\sigma_b^2+\sigma_w^2/m_i)$ and
$(m_i-1)S_i^2\sim\sigma_w^2\chi_{m_i-1}^2$. 
Here $\{\bar{Y}_i:i=1,\cdots,n\}$ and $\{S_i^2::i=1,\cdots,n\}$ are essentially the sample means and variances, 
and hence they are independent random variables.

Denote $SSE=\sum_{i=1}^n (m_i-1)S_i^2$. We have $SSE\sim\sigma_w^2\chi_{N-n}^2$, where $N=\sum_{i=1}^nm_i$.
Let $sse$ denote the value of $SSE$ based on the observed data. 
Define 
\begin{equation}
Q_w= \frac{sse}{SSE/\sigma_w^2}.
\end{equation}
Note that given the observed data, $Q_w$ follows a scaled inverse chi-square distribution,
and furthermore, $Q_w$ equals to $\sigma_w^2$ at the observed data, i.e. when $SSE$ takes value $sse$.

Denote $W_i = 1/(\sigma_b^2+\sigma_w^2/m_i)$, $i=1,\cdots,n$. 
Let $\bar{Y} = \sum_{i=1}^n W_i\bar{Y}_i/(\sum_{i=1}^n W_i)$.
We note that $\bar{Y}_i\sim\MN(\mu,W_i^{-1})$, and
$\bar{Y}\sim\MN(\mu, 1/(\sum_{i=1}^n W_i))$.
Treating $\bar{Y}_i$ as the outcomes regressed onto a constant with weight $W_i$, 
and based on the standard linear regression theory, we can check that
\begin{equation}
 SSR=\sum_{i=1}^n W_i(\bar{Y}_i-\bar{Y})^2
  = \sum_{i=1}^n\frac{\bar{Y}_i^2}{\sigma_b^2+\sigma_w^2/m_i}
   - \frac{[\sum_{i=1}^n\bar{Y}_i/(\sigma_b^2+\sigma_w^2/m_i)]^2}{\sum_{i=1}^n(\sigma_b^2+\sigma_w^2/m_i)^{-1}}, 
  \label{SSR}
\end{equation}
is a $\chi_{n-1}^2$ random variable.
Furthermore $\bar{Y}$ and $SSR$ are independent.
Based on equation~(\ref{SSR}),
we can solve $\sigma_b^2$ as a deterministic function of $(\bar{\bY},\sigma_w^2,SSR)$,
denoted as $h(\bar{\bY},\sigma_w^2,SSR)$.
Here $\bar{\bY}=(\bar{Y}_1,\cdots,\bar{Y}_n)$.
Define a generalized test statistic
\begin{equation}
 Q_b=h(\bar{\by},Q_w,SSR),
\end{equation}
where $\bar{\by}=(\bar{y}_1,\cdots,\bar{y}_n)$ denote the observed sample means.
Given the observed data, both the distributions of $Q_w$ and $SSR$ are known,
hence $Q_b$ has a distribution free of any parameters, and 
$Q_b$ equals to $\sigma_b^2$ at the observed data.

Denote $Z=(\bar{Y}-\mu)\sqrt{\sum_{i=1}^n W_i}$. We have $Z\sim\MN(0,1)$.
Therefore, we can write $\mu=\bar{Y} - Z/\sqrt{\sum_{i=1}^n W_i}$ \citep{MuCI2004}.
Define the following generalized test statistic
\begin{equation}
Q_{\mu} = \bigg[ \tilde{y} - Z\Big(\sum_{i=1}^n\tilde{W}_i\Big)^{-1/2} \bigg]^2, \quad
\tilde{W}_i = \frac{1}{Q_b+Q_w/m_i}, \quad
\tilde{y} = \frac{\sum_{i=1}^n \bar{y}_i\tilde{W}_i}{\sum_{i=1}^n\tilde{W}_i}.
\end{equation}
Given the observed data,
$Q_{\mu}$ has a known distribution free of any parameters.
Furthermore, at the observed data, $Q_b$ and $Q_w$ reduce to $\sigma_b^2$ and $\sigma_w^2$,
and hence $Q_{\mu}$ equals $\mu^2$.

\subsection{Generalized test p-value and generalized CI calculation}

For inference on the RMS parameter $\rho$, we consider the following generalized test statistic,
\begin{equation}
Q = Q_w+Q_b+Q_{\mu}.
\end{equation}
We treat those observed quantities, e.g. $(\bar{\by},sse)$, as constants, and the
distribution of $Q$ is induced by $(SSE,SSR,Z)$.
We note that the generalized test statistic $Q$ and its distribution require only
the summary statistics: $sse$ and $\bar{\by}$.
It is clear that the distribution of $Q$ does not depend on any unknown parameters given the observed data.
We also note that for the observed data, $Q$ reduces to $\rho^2=\sigma_w^2+\sigma_b^2+\mu^2$. 
The generalized p-value for testing $H_0:\rho\geq\rho_0$ can be computed as $\Pr(Q\geq\rho_0^2)$, 
and a $100(1-\alpha)\%$ generalized CI for $\rho$ can be computed as $[\sqrt{Q_{(\alpha/2)}}, \sqrt{Q_{(1-\alpha/2)}}]$,
where $Q_{(\alpha)}$ denotes the $\alpha$-th percentile of the distribution of $Q$
\citep{GP1989,GCI1993}.

Following the generalized test literature,
the statistical inference can be made based on generating random numbers from the distribution of $Q$ as follows:
\begin{enumerate}
\item[(i)] simulate $(SSE/\sigma_w^2)$ from the $\chi_{N-n}^2$ distribution to generate $Q_w$;
\item[(ii)] simulate $SSR$ from the $\chi_{n-1}^2$ distribution, and compute $Q_b$ from $Q_w$ and $SSR$;
\item[(iii)] simulate $Z$ from $\MN(0,1)$ distribution, and compute $Q_{\mu}$ from $Q_w$, $Q_b$ and $Z$;
\item[(iv)] compute $Q=Q_w+Q_b+Q_{\mu}$.
\end{enumerate}

Given the simulated random realizations, denoted as $(Q^1,\cdots,Q^B)$,
we compute the generalized test p-value for testing $H_0:\rho\geq \rho_0$ 
\begin{equation}
 P=\frac{\sum_{i=1}^BI(Q^i\ge \rho_0^2)}{B}, 
\end{equation}
and compute the
generalized CI for $\rho$ with $1-\alpha$ coefficient as 
\begin{equation}
 \Big[ \sqrt{Q^{(B\alpha/2)}}, \sqrt{Q^{(B(1-\alpha/2))}} \Big],
\end{equation}
where $Q^{(k)}$ denotes the ordered Q-values. 

\subsection{Numerical computation}
Note that conditional on the $Q_w$ and $Q_b$, 
$\sqrt{Q_{\mu}}$ follows a normal distribution with mean $\tilde{y}$ and variance
$1/(\sum_i\tilde{W}_i)$. Hence we can easily compute
$\Pr(Q\geq\rho_0^2|Q_w,Q_b) = S_{\chi_1^2}[(\sum_i\tilde{W}_i)(\rho_0^2-Q_w-Q_b)|\tilde{y}^2(\sum_i\tilde{W}_i)]$,
where $S_{\chi_1^2}(\cdot|\lambda)$ denotes the tail probability of a $\chi_1^2$ distribution
with non-centrality parameter $\lambda$.
Therefore we can compute the generalized test p-value as
\[
P = \frac{1}{B}\sum_{k=1}^B \Pr(Q\geq\rho_0^2|Q_w^k,Q_b^k).
\]
The corresponding generalized CI can be computed by reversely solving the critical values of $Q$
at the generalized p-values of $\alpha/2$ and $(1-\alpha/2)$.
We note that this practice amounts to analytically integrating out the random variable
$Z$ from the marginal distribution of $Q$, which avoids the need for actual Monte Carlo
simulation of $Z$ and can lead to more accurate test p-value and CI calculation.

\subsection{Large-sample normal approximation}
The existing large-sample test methods 
are based on the normal approximation of $R=\sum_{i,j}Y_{ij}^2/N$
with mean and variance \citep{Pennello2002,Pennello2003,ndikintum2016}
\begin{equation}
\E(R)=\rho^2, \quad
Var(R) = \frac{2}{N^2}\sum_{i=1}^n(\sigma_w^2+m_i\sigma_b^2)^2+(m_i-1)\sigma_w^4
+2m_i(\sigma_w^2+m_i\sigma_b^2)\mu^2.
\end{equation}
The parameter estimates can be readily obtained from fitting a LMM to the data.
The existing methods differ in their calculation of the variance,
leading to a score test and Wald test, denoted as Z-score and Z-Wald respectively,
depending on whether the variance is computed under the null hypothesis or not.

In the next section we conduct simulation studies to investigate the performance
of proposed method (with $B=10^4$, denoted as GT) compared to the 
existing large-sample test methods. 

\section{Numerical studies}

In the numerical studies, we have found that the Z-Wald test has very liberal
type I errors (the complete results are available at the Supporting Information Section S1). Hence we only include
the Z-score test in the following comparison.

We conducted two sets of simulations. First, we simulate an unbalanced study with parameters,
$(\mu=-0.57,\sigma_w=1.48, \sigma_b=1.38)$, estimated from
a pulse oximetry comparison study (see Section 4). We consider 
$n=16$ subjects with $m_i$ ranging from 5 to 20, and $n=20$ subjects
with $m_i$ ranging from 5 to 24.
We set $\rho_0=2.1$ under the null hypothesis and $\rho_0=3$ for estimating power.
Secondly, we consider a balanced study with 
combinations of $n=(10,20,30)$, $m=(5,10,20)$, 
$\sigma_w^2+\sigma_b^2=(0.2,0.4,0.8)\times\rho^2$, $\sigma_w^2/\sigma_b^2=(1/3,1,3)$, and 
setting $\mu=\sqrt{\rho^2-\sigma_w^2-\sigma_b^2}$.
We test $H_0: \rho\geq 3$, and set $\rho=3$ under the null hypothesis and $\rho=\sqrt{6}$ for estimating power.

\subsection{Type I errors}

We conduct $10^4$ null simulations to evaluate the type I errors at
the significance level $\alpha=$ 0.01 and 0.05. 
Overall we have obtained very similar patterns across all simulation scenarios. Here
we present the results for the unbalanced study and 
the balanced study with $n=20,m=10$. The complete simulation
results can be found in the Supporting Information Section S1.

Table~\ref{err1} summarizes the empirical type I errors for the unbalanced study with
$n=16,20$ subjects. 
Table~\ref{err2}
shows the empirical type I errors for balanced study with
$(\sigma_w^2+\sigma_b^2)/\rho^2=(0.2,0.4,0.8)$
and $\sigma_w^2/\sigma_b^2=(1/3,1,3)$ at $\alpha=0.05,0.01$.
The proposed test (denoted as GT) is compared to the Z-score test.
Overall the results show that the type I errors are well controlled for the
proposed method, while the Z-score test is more conservative especially at more
stringent significance level. For the proposed GT, generally
larger $\mu$ (smaller $\sigma_w^2+\sigma_b^2$) leads to more accurate control 
of type I errors.
\begin{table}[h]
\caption{Empirical type I errors at significance level $\alpha$ estimated over $10^4$ simulations.
GT is the proposed test, Z-score is the score based Z-test.
Data are simulated from $n$ subjects with 
parameters $(\mu=-0.57,\sigma_w=1.48, \sigma_b=1.38)$. }
\centering
\label{err}
\begin{adjustbox}{max width=1\textwidth}
\begin{tabular}{c|cc|cc} 
\hline
   & \multicolumn{2}{c|}{$n=16$} &  \multicolumn{2}{c}{$n=20$}     \\
   $\alpha$    & 0.05 & 0.01 & 0.05 & 0.01 \\
  \hline
  GT    & 0.028 & 0.005   & 0.032 & 0.006  \\
  Z-score  & 0.023 & 0.0003 &  0.026 & 0.0004  \\ 
\hline
\end{tabular}
\label{err1}
\end{adjustbox}
\end{table}

\begin{table}[h]
\caption{Empirical type I errors at significance level $\alpha$ estimated over $10^4$ null simulations.
Data are simulated from $n=20$ subjects each with $m=10$ measures and $\rho=3$.
GT is the proposed test, Z-score is the score based Z-test. }
\centering
\label{err2}
\begin{adjustbox}{max width=1\textwidth}
\begin{tabular}{c|ccc|ccc|ccc} 
\hline
  $(\sigma_w^2+\sigma_b^2)/\rho^2$   &  \multicolumn{3}{c|}{0.2}   &  \multicolumn{3}{c|}{0.4}  
  &  \multicolumn{3}{c}{0.8}      \\
\hline
 $\sigma_w^2:\sigma_b^2$     &  
    1:3 & 1:1 & 3:1 &  1:3 & 1:1 & 3:1 & 1:3 & 1:1 & 3:1   \\
  \hline
     \multicolumn{10}{c}{$\alpha=0.05$}      \\
\hline
  GT     & 0.041 & 0.047  & 0.045  & 0.040  & 0.045  & 0.039  & 0.030 & 0.034  
          & 0.033  \\ 
  Z-score & 0.034 & 0.041  & 0.039 & 0.033  & 0.039 & 0.036 & 0.026 & 0.032 
          & 0.032  \\ 
\hline
     \multicolumn{10}{c}{$\alpha=0.01$}      \\
\hline
  GT     & 0.0073 & 0.0079  & 0.0100  & 0.0083  & 0.0090  & 0.0077  & 0.0056 & 0.0063  
          & 0.0059  \\ 
  Z-score & 0.0019 & 0.0033  & 0.0050 & 0.0019  & 0.0024 & 0.0026 & 0.0007 & 0.0009 
          & 0.0027  \\ 
\hline
\end{tabular}
\end{adjustbox}
\end{table}

\subsection{Power}

For the unbalanced study, we consider testing $H_0: \rho>3$.
For the balanced study, we set $\rho_0=\sqrt{6}$ and consider testing $H_0: \rho>3$. 
We use $10^4$ Monte Carlo simulations to estimate power under each configuration. 
Overall, we have observed very similar conclusions across simulation scenarios.
Here we report the results for the unbalanced study and balanced study with $n=20,m=10$.
The complete results are available at the Supporting Information Section S1.

Table~\ref{pwr1} summarizes the power for the unbalanced study.
Table~\ref{pwr2} summarizes the power for 
$(\sigma_w^2+\sigma_b^2)/6=(0.2,0.8)$
and $\sigma_w^2/\sigma_b^2=(1/3,3)$ under the 0.01 and 0.05 significance levels.
Overall the proposed method
performs remarkably better than the Z-score test by a large margin, especially at more stringent significance level. 
For the proposed GT,
(1) smaller between-subject variance $\sigma_b^2$
leads to higher power;
and
(2) larger $\mu$ (smaller $\sigma_w^2+\sigma_b^2$) also leads to higher power.
\begin{table}[h] 
\caption{Power (\%) at significance level $\alpha$ estimated over $10^4$ simulations.
GT is the proposed test, Z-score is the score based Z-test.
Data are simulated from $n$ subjects with 
parameters $(\mu=-0.57,\sigma_w=1.48, \sigma_b=1.38)$. 
We are testing $H_0:\rho\geq 3$. }
\centering
\begin{adjustbox}{max width=1\textwidth}
\begin{tabular}{c|cc|cc} 
\hline
   & \multicolumn{2}{c|}{$n=16$} &  \multicolumn{2}{c}{$n=20$}     \\
   $\alpha$    & 0.05 & 0.01 & 0.05 & 0.01 \\
  \hline
  GT    & 82.1 & 49.3   & 92.7 & 68.2  \\
  Z-score  & 72.1 & 3.2 &  86.2 & 16.6  \\ 
\hline
\end{tabular}
\label{pwr1}
\end{adjustbox}
\end{table}

\begin{table}[h]
\caption{Power (\%) at significance level $\alpha$ estimated over $10^4$ simulations.
Data are simulated for $n=20$ subjects each with $m=10$ measures under $\sigma_w^2+\sigma_b^2+\mu^2=6$.
We are testing $H_0: \rho\geq 3$.
GT is the proposed test, and Z-score is the score based Z-test. }
\centering
\label{pwr2}
\begin{adjustbox}{max width=1\textwidth}
\begin{tabular}{c|ccc|ccc|ccc} 
\hline
  $(\sigma_w^2+\sigma_b^2)/\rho^2$   &  \multicolumn{3}{c|}{0.2}   &  \multicolumn{3}{c|}{0.4}  
  &  \multicolumn{3}{c}{0.8}      \\
\hline
 $\sigma_w^2:\sigma_b^2$     &  
    1:3 & 1:1 & 3:1 &  1:3 & 1:1 & 3:1 & 1:3 & 1:1 & 3:1   \\
  \hline
     \multicolumn{10}{c}{$\alpha=0.05$}      \\
\hline
  GT     & 75.4 & 88.0  & 98.2  & 51.2  & 66.2  & 87.8  & 33.0 & 48.1  
          & 73.2  \\ 
  Z-score & 72.1 & 85.9  & 97.7 & 46.5  & 62.2 & 86.2 & 30.5 & 45.0 
          & 71.6  \\ 
\hline
     \multicolumn{10}{c}{$\alpha=0.01$}      \\
\hline
  GT     & 44.7 & 62.8  & 89.0  & 21.2  & 34.1  & 61.4  & 11.0 & 19.5
          & 41.0  \\ 
  Z-score & 25.9 & 43.4  & 78.5 & 7.6  & 16.2 & 41.7 & 2.4 & 5.5 
          & 18.9  \\ 
\hline
\end{tabular}
\end{adjustbox}
\end{table}

\subsection{CI calculation}
Table~\ref{CI1} and \ref{CI2} summarize the coverage probability (CP)
and average width (AW) of computed 90\% CIs over $10^4$ simulations. 
Overall we can see that the CIs produced by the proposed method (denoted as GCI) 
have very good coverage probability close to the nominal 90\% level.
As expected, the Z-score test based approach has more conservative
performance, with CP generally much larger than the desired 90\% level and
CI much wider than the GCI. 
\begin{table}[h]
\caption{Coverage probability (CP) and average width (AW) of 90\% CI for $\rho$ estimated over $10^4$ simulations.
GCI is the proposed test, Z-score is the score based Z-test.
Data are simulated from an unbalanced study with $n$ subjects and $\rho=2.1$. }
\centering
\label{err}
\begin{adjustbox}{max width=1\textwidth}
\begin{tabular}{c|cc|cc} 
\hline
   & \multicolumn{2}{c|}{$n=16$} &  \multicolumn{2}{c}{$n=20$}     \\
       & CP & AW & CP & AW \\
  \hline
  GCI    & 0.892 & 0.815   & 0.891 & 0.693  \\
  Z-score  & 0.973 & 0.836 &  0.965 & 0.728  \\ 
\hline
\end{tabular}
\label{CI1}
\end{adjustbox}
\end{table}
\begin{table}[h]
\caption{Coverage probability (CP) and average width (AW) of 90\% CI estimated over $10^4$ simulations.
Data are simulated from a balanced study with $n=20$ subjects each with $m=10$ measures and $\rho=3$.
GCI is the proposed test, Z-score is the score based Z-test. }
\centering
\label{err}
\begin{adjustbox}{max width=1\textwidth}
\begin{tabular}{c|ccc|ccc|ccc} 
\hline
  $(\sigma_w^2+\sigma_b^2)/\rho^2$   &  \multicolumn{3}{c|}{0.2}   &  \multicolumn{3}{c|}{0.4}  
  &  \multicolumn{3}{c}{0.8}      \\
\hline
 $\sigma_w^2:\sigma_b^2$     &  
    1:3 & 1:1 & 3:1 &  1:3 & 1:1 & 3:1 & 1:3 & 1:1 & 3:1   \\
  \hline
     \multicolumn{10}{c}{CP}      \\
\hline
  GCI     & 0.899 & 0.898  & 0.902  & 0.900 & 0.896  & 0.895  & 0.890 & 0.892  
          & 0.896  \\ 
  Z-score & 0.933 & 0.929  & 0.923 & 0.951  & 0.939 & 0.929 & 0.963 & 0.963 
          & 0.953  \\ 
\hline
     \multicolumn{10}{c}{AW}      \\
\hline
  GCI     & 0.841 & 0.701  & 0.535  & 1.109  & 0.909  & 0.687  & 1.357 & 1.055  
          & 0.761  \\ 
  Z-score & 0.874 & 0.720  & 0.539 & 1.187  & 0.956 & 0.707 & 1.450 & 1.091 
          & 0.772  \\ 
\hline
\end{tabular}
\label{CI2}
\end{adjustbox}
\end{table}

\section{Application to oximetry comparison study}

For illustration, we consider an oximetry comparison study conducted by a medical device company 
to demonstrate equivalency in performance between an FDA-cleared oximetry system 
and an investigational oximeter.
The study has obtained multiple measures of difference of tissue oxygen saturation levels
for a cohort of healthy, non-smoking adults and adolescent volunteers.
We model the measured oxygen differences with a linear mixed-effects model (LMM),
and investigate the system equivalency  based on testing $H_0: \rho\geq 3\%$.

For illustrative purposes, we analyze a representative dataset
from 16 individuals and provide the summary data. 
Table~\ref{poData} listed the sample size $m_i$ and mean $\bar{y}_i$ for each individual.
The observed $sse=221.037\%^2$.
We note that the LMM estimation requires only $(\bar{\by},sse)$ without the need of individual level data, 
similar to the proposed generalized test method 
(see Appendix for details).

\begin{table}[h]
\centering
\caption{Individual sample sizes and means (\%) for the oximetry comparison study}
\begin{tabular}{ccc|ccc}
  \hline
  Individual $i$ & $m_i$ & $\bar{y}_i$ & Individual $i$ & $m_i$ & $\bar{y}_i$  \\
  \hline
1 &    9 & -0.026  &    9 &   10 & 0.963  \\
2 &   10 & 0.447  &   10 &   10 & 0.643  \\
3 &   10 & 0.083  &   11 &   10 & -0.200  \\
4 &   10 & -0.103  &   12 &   10 & -1.337  \\
5 &    5 & -2.587  &   13 &    2 & -4.333  \\
6 &   10 & -0.610  &   14 &   10 & -2.807  \\
7 &   10 & 0.040  &   15 &   10 & 0.563  \\
8 &   10 & -0.593  &   16 &   10 & -0.797  \\
  \hline
\end{tabular}
\label{poData}
\end{table}

\medskip
The proposed GT (using $B=10^4$ simulations) yields a significant p-value of 0.006, and the
computed 90\% GCI for the RMS parameter $\rho$ is [1.665,2.528]\%. 
The Z-score test reported a p-value of 0.010
with computed 90\% CI for $\rho^2$ as [-1.447, 7.221]$\%^2$,
leading to a 90\% CI for $\rho$ as [0,2.687]\% for $\rho$. 
Both tests supported equivalency of the two systems.
The proposed test produced a more significant test p-value and a much narrower CI,
offering stronger support for the system equivalency.

\section{Discussion}

We have proposed a generalized significance test method with accurate CI calculation for the root mean squares parameter (RMS) in a
linear mixed-effects model. The RMS parameter directly quantifies the total absolute variation of outcomes and
can be used to test equivalency in paired medical device comparison studies. 
Compared to the existing large-sample test methods, our proposed
method shows more powerful performance and produces CIs with more accurate coverage probabilities. 
When applied to a pulse oximetry equivalency study, our proposed method
yielded more signifiant test p-value
and stronger evidence to  support the system equivalency, as compared to the existing methods.
We have also conducted extensive numerical studies to illustrate the very favorable performance of our proposed method. 
The new approach has been implemented in a publicly available R package.
The new method along with the R package provide useful and practical tools,
and timely bridge an existing gap in the field.

\section{Acknowledgments}

This research was supported in part by NIH grant GM083345 and CA134848, and NSF grant DMS-1902903.
We are grateful to the University of Minnesota Supercomputing Institute for 
assistance with the computations.

\section*{Supporting information}

An R package implementing the new method is 
available at \url{http://github.com/baolinwu/RAMgt}.
Supporting information contains complete simulation results (Section S1) 
and sample scripts to install and use the
developed R package (Section S2).

\bibliographystyle{biom}
\bibliography{GCI}

\appendix

\section{Sufficient statistics, likelihood calculation, and parameter estimation}

Denote $\Sigma_i=Cov(\bY_i)$ for $i=1,\cdots,n$.
We can write  $\Sigma_i=\sigma_b^2\bo_{m_i}\bo_{m_i}^T+\sigma_w^2\bI_{m_i}$, with $\bo_{m_i}$ being
a column vector of ones and $\bI_{m_i}$ an $m_i$-th oder identity matrix.
Noting that $\bo_{m_i}\bo_{m_i}^T/m_i$ is 
a rank-one projection matrix, $\Sigma_i$ has two unique eigenvalues:
$\sigma_w^2$ (with algebraic multiplicity $m_i-1$) and $\sigma_w^2+m_i\sigma_b^2$ (with algebraic multiplicity 1). 
We can analytically compute its inverse and determinant,
and check that $-2\log[\Pr(\by_i)]\propto \log(\sigma_w^2+m_i\sigma_b^2)+(m_i-1)\log(\sigma_w^2)
           + \sigma_w^{-2}\|\by_i-\bar{y}_i\|^2   
           + (\sigma_w^2+m_i\sigma_b^2)^{-1}m_i(\bar{y}_i-\mu)^2$ 
           (up to some constant that does not depend on data).
The log likelihood can then be shown proportional to
\[
 (N-n)\log(\sigma_w^2) + \frac{sse}{\sigma_w^{2}}
 + \sum_{i=1}^n \Big\{ \log(\sigma_w^2+m_i\sigma_b^2)+\frac{m_i(\bar{y}_i-\mu)^2}{\sigma_w^2+m_i\sigma_b^2} \Big\},
\]
where $sse = \sum_i\|\by_i-\bar{y}_i\|^2$. 
Therefore, we just need $(\bar{\by},sse)$  
to compute likelihood to estimate parameters, the same as the proposed generalized test method,
as shown previously.

\clearpage

\renewcommand{\thesection}{S\arabic{section}}   
\renewcommand{\thetable}{S\arabic{table}}   
\renewcommand{\thefigure}{S\arabic{figure}}

\setcounter{section}{0}
\setcounter{table}{0}

\begin{centering}
{\Large \bf Supporting Information }
\end{centering}

\section{Simulation study}

We conducted two sets of simulations. First, we simulate an unbalanced study with parameters,
$(\mu=-0.57,\sigma_w=1.48, \sigma_b=1.38)$, estimated from
a pulse oximetry comparison study (see main text Section 4). We consider 
$n=16$ subjects with $m_i$ ranging from 5 to 20, and $n=20$ subjects
with $m_i$ ranging from 5 to 24.
We set $\rho_0=2.1$ under the null hypothesis and $\rho_0=3$ for estimating power.
Secondly, we consider a balanced study with 
combinations of $n=(10,20,30)$, $m=(5,10,20)$, 
$\sigma_w^2+\sigma_b^2=(0.2,0.4,0.8)\times\rho^2$, $\sigma_w^2/\sigma_b^2=(1/3,1,3)$, and 
setting $\mu=\sqrt{\rho^2-\sigma_w^2-\sigma_b^2}$.
We test $H_0: \rho\geq 3$, and set $\rho=3$ under the null hypothesis and $\rho=\sqrt{6}$ for estimating power.
The proposed test (denoted as GT) is compared to the Z-score/Z-Wald tests.

The main text contains the simulation results for the unbalanced studies.
Here we summarize the complete simulation results for the balanced studies.

\subsection{Type I errors}

We conduct $10^4$ null simulations to evaluate the type I errors at
the significance level $\alpha=$ 0.01 and 0.05. 
Tables S1-S3 summarize the type I errors for the balanced studies.
Overall we have obtained very similar patterns across all simulation scenarios. 
The type I errors are well controlled for the proposed method. 
The Z-score test is more conservative especially at more
stringent significance level, and the Z-Wald test has severely inflated type I errors.
The Z-score test has largely improved performance with increasing sample sizes.
In contrast, the GT has relatively stable performance across all scenarios.
Overall, larger $\mu$ (smaller $\sigma_w^2+\sigma_b^2$) leads to more accurate control 
of type I errors for the proposed GT.
\begin{table}[h]
\caption{Empirical type I errors at significance level $\alpha$ estimated over $10^4$ null simulations.
Data are simulated from $n=10$ subjects each with $m$ measures and $\rho=3$.
GT is the proposed test, Z-score is the score based Z-test. }
\centering
\label{err}
\begin{adjustbox}{max width=1\textwidth}
\begin{tabular}{c|ccc|ccc|ccc} 
\hline
    \multicolumn{10}{c}{$n=10,m=5$}      \\
\hline
  $(\sigma_w^2+\sigma_b^2)/\rho^2$   &  \multicolumn{3}{c|}{0.2}   &  \multicolumn{3}{c|}{0.4}  
  &  \multicolumn{3}{c}{0.8}      \\
\hline
 $\sigma_w^2:\sigma_b^2$     &  
    1:3 & 1:1 & 3:1 &  1:3 & 1:1 & 3:1 & 1:3 & 1:1 & 3:1   \\
  \hline
     \multicolumn{10}{c}{$\alpha=0.05$}      \\
\hline
  GT     & 0.043 & 0.043  & 0.040  & 0.040  & 0.037  & 0.030  & 0.026 & 0.024  
          & 0.017  \\ 
  Z-score & 0.027 & 0.031  & 0.034 & 0.024  & 0.026 & 0.026 & 0.016 & 0.021 
          & 0.025  \\ 
  Z-Wald & 0.110 & 0.106  & 0.096 & 0.137  & 0.124 & 0.110 & 0.143 & 0.136 
          & 0.115  \\ 
\hline
     \multicolumn{10}{c}{$\alpha=0.01$}      \\
\hline
  GT     & 0.0082 & 0.0082  & 0.0063  & 0.0072  & 0.0076  & 0.0045  & 0.0048 & 0.0043  
          & 0.0025  \\ 
  Z-score & 0.0001 & 0.0002  & 0.0007 & 0  & 0.0001 & 0.0003 & 0 & 0.0001 
          & 0.0002  \\ 
  Z-Wald & 0.056 & 0.050  & 0.042 & 0.072  & 0.066 & 0.052 & 0.083 & 0.078 
          & 0.054  \\ 
\hline
    \multicolumn{10}{c}{$n=10,m=10$}      \\
\hline
  $(\sigma_w^2+\sigma_b^2)/\rho^2$   &  \multicolumn{3}{c|}{0.2}   &  \multicolumn{3}{c|}{0.4}  
  &  \multicolumn{3}{c}{0.8}      \\
\hline
 $\sigma_w^2:\sigma_b^2$     &  
    1:3 & 1:1 & 3:1 &  1:3 & 1:1 & 3:1 & 1:3 & 1:1 & 3:1   \\
  \hline
     \multicolumn{10}{c}{$\alpha=0.05$}      \\
\hline
  GT     & 0.042 & 0.040  & 0.041  & 0.036  & 0.038  & 0.036  & 0.026 & 0.025  
          & 0.024  \\ 
  Z-score & 0.029 & 0.027  & 0.030 & 0.022  & 0.023 & 0.029 & 0.016 & 0.017 
          & 0.025  \\ 
  Z-Wald & 0.109 & 0.102  & 0.093 & 0.122  & 0.123 & 0.104 & 0.153 & 0.142 
          & 0.116  \\ 
\hline
     \multicolumn{10}{c}{$\alpha=0.01$}      \\
\hline
  GT     & 0.0083 & 0.0068  & 0.0072  & 0.0075  & 0.0079  & 0.0074  & 0.0050 & 0.0038  
          & 0.0027  \\ 
  Z-score & 0.0006 & 0.0001  & 0.0007 & 0  & 0 & 0.0004 & 0 & 0 
          & 0.0003  \\ 
  Z-Wald & 0.051 & 0.046  & 0.039 & 0.066  & 0.063 & 0.049 & 0.085 & 0.078 
          & 0.056  \\ 
\hline
    \multicolumn{10}{c}{$n=10,m=20$}      \\
\hline
  $(\sigma_w^2+\sigma_b^2)/\rho^2$   &  \multicolumn{3}{c|}{0.2}   &  \multicolumn{3}{c|}{0.4}  
  &  \multicolumn{3}{c}{0.8}      \\
\hline
 $\sigma_w^2:\sigma_b^2$     &  
    1:3 & 1:1 & 3:1 &  1:3 & 1:1 & 3:1 & 1:3 & 1:1 & 3:1   \\
  \hline
     \multicolumn{10}{c}{$\alpha=0.05$}      \\
\hline
  GT     & 0.043 & 0.044  & 0.046  & 0.036  & 0.042  & 0.039  & 0.030 & 0.027  
          & 0.029  \\ 
  Z-score & 0.027 & 0.030  & 0.033 & 0.023  & 0.027 & 0.027 & 0.017 & 0.018 
          & 0.023  \\ 
  Z-Wald & 0.110 & 0.105  & 0.094 & 0.131  & 0.127 & 0.105 & 0.157 & 0.141 
          & 0.124  \\ 
\hline
     \multicolumn{10}{c}{$\alpha=0.01$}      \\
\hline
  GT     & 0.0074 & 0.0085  & 0.0091  & 0.0082  & 0.0079  & 0.0078  & 0.0057 & 0.0060  
          & 0.0052  \\ 
  Z-score & 0 & 0.0006  & 0.0002 & 0  & 0 & 0.0001 & 0 & 0.0001 
          & 0.0001  \\ 
  Z-Wald & 0.055 & 0.047  & 0.041 & 0.068  & 0.065 & 0.048 & 0.094 & 0.081 
          & 0.061  \\ 
\hline
\end{tabular}
\end{adjustbox}
\end{table}

\begin{table}[h]
\caption{Empirical type I errors at significance level $\alpha$ estimated over $10^4$ null simulations.
Data are simulated from $n=20$ subjects each with $m$ measures and $\rho=3$.
GT is the proposed test, Z-score is the score based Z-test. }
\centering
\label{err}
\begin{adjustbox}{max width=1\textwidth}
\begin{tabular}{c|ccc|ccc|ccc} 
\hline
    \multicolumn{10}{c}{$n=20,m=5$}      \\
\hline
  $(\sigma_w^2+\sigma_b^2)/\rho^2$   &  \multicolumn{3}{c|}{0.2}   &  \multicolumn{3}{c|}{0.4}  
  &  \multicolumn{3}{c}{0.8}      \\
\hline
 $\sigma_w^2:\sigma_b^2$     &  
    1:3 & 1:1 & 3:1 &  1:3 & 1:1 & 3:1 & 1:3 & 1:1 & 3:1   \\
  \hline
     \multicolumn{10}{c}{$\alpha=0.05$}      \\
\hline
  GT     & 0.046 & 0.044  & 0.045  & 0.043  & 0.042  & 0.040  & 0.033 & 0.030  
          & 0.029  \\ 
  Z-score & 0.038 & 0.038  & 0.042 & 0.033  & 0.036 & 0.039 & 0.031 & 0.031 
          & 0.037  \\ 
  Z-Wald & 0.091 & 0.085  & 0.081 & 0.103  & 0.097 & 0.090 & 0.120 & 0.111 
          & 0.101  \\ 
\hline
     \multicolumn{10}{c}{$\alpha=0.01$}      \\
\hline
  GT     & 0.0079 & 0.0086  & 0.0089  & 0.0079  & 0.0078  & 0.0078  & 0.0055 & 0.0057  
          & 0.0039  \\ 
  Z-score & 0.0029 & 0.0046  & 0.0044 & 0.0019  & 0.0020 & 0.0026 & 0.0010 & 0.0008 
          & 0.0019  \\ 
  Z-Wald & 0.037 & 0.033  & 0.031 & 0.047  & 0.043 & 0.037 & 0.060 & 0.054 
          & 0.043  \\ 
\hline
    \multicolumn{10}{c}{$n=20,m=10$}      \\
\hline
  $(\sigma_w^2+\sigma_b^2)/\rho^2$   &  \multicolumn{3}{c|}{0.2}   &  \multicolumn{3}{c|}{0.4}  
  &  \multicolumn{3}{c}{0.8}      \\
\hline
 $\sigma_w^2:\sigma_b^2$     &  
    1:3 & 1:1 & 3:1 &  1:3 & 1:1 & 3:1 & 1:3 & 1:1 & 3:1   \\
  \hline
     \multicolumn{10}{c}{$\alpha=0.05$}      \\
\hline
  GT     & 0.041 & 0.047  & 0.045  & 0.040  & 0.045  & 0.039  & 0.030 & 0.034  
          & 0.033  \\ 
  Z-score & 0.034 & 0.041  & 0.039 & 0.033  & 0.039 & 0.036 & 0.026 & 0.032 
          & 0.032  \\ 
  Z-Wald & 0.084 & 0.089  & 0.079 & 0.103  & 0.099 & 0.082 & 0.118 & 0.118 
          & 0.099  \\ 
\hline
     \multicolumn{10}{c}{$\alpha=0.01$}      \\
\hline
  GT     & 0.0073 & 0.0079  & 0.0100  & 0.0083  & 0.0090  & 0.0077  & 0.0056 & 0.0063  
          & 0.0059  \\ 
  Z-score & 0.0019 & 0.0033  & 0.0050 & 0.0019  & 0.0024 & 0.0026 & 0.0007 & 0.0009 
          & 0.0027  \\ 
  Z-Wald & 0.032 & 0.033  & 0.028 & 0.044  & 0.045 & 0.033 & 0.057 & 0.056 
          & 0.041  \\ 
\hline
    \multicolumn{10}{c}{$n=20,m=20$}      \\
\hline
  $(\sigma_w^2+\sigma_b^2)/\rho^2$   &  \multicolumn{3}{c|}{0.2}   &  \multicolumn{3}{c|}{0.4}  
  &  \multicolumn{3}{c}{0.8}      \\
\hline
 $\sigma_w^2:\sigma_b^2$     &  
    1:3 & 1:1 & 3:1 &  1:3 & 1:1 & 3:1 & 1:3 & 1:1 & 3:1   \\
  \hline
     \multicolumn{10}{c}{$\alpha=0.05$}      \\
\hline
  GT     & 0.047 & 0.047  & 0.044  & 0.042  & 0.044  & 0.045  & 0.034 & 0.034  
          & 0.035  \\ 
  Z-score & 0.040 & 0.041  & 0.038 & 0.034  & 0.037 & 0.040 & 0.031 & 0.029 
          & 0.035  \\ 
  Z-Wald & 0.090 & 0.084  & 0.076 & 0.100  & 0.100 & 0.0915 & 0.120 & 0.117 
          & 0.105  \\ 
\hline
     \multicolumn{10}{c}{$\alpha=0.01$}      \\
\hline
  GT     & 0.0090 & 0.0092  & 0.0095  & 0.0077  & 0.0073  & 0.0100  & 0.0066 & 0.0052  
          & 0.0063  \\ 
  Z-score & 0.0032 & 0.0029  & 0.0043 & 0.0017  & 0.0022 & 0.0039 & 0.0007 & 0.0005 
          & 0.0020  \\ 
  Z-Wald & 0.038 & 0.032  & 0.025 & 0.045  & 0.042 & 0.035 & 0.064 & 0.057 
          & 0.046  \\ 
\hline
\end{tabular}
\end{adjustbox}
\end{table}

\begin{table}[h]
\caption{Empirical type I errors at significance level $\alpha$ estimated over $10^4$ null simulations.
Data are simulated from $n=30$ subjects each with $m$ measures and $\rho=3$.
GT is the proposed test, Z-score is the score based Z-test. }
\centering
\label{err}
\begin{adjustbox}{max width=1\textwidth}
\begin{tabular}{c|ccc|ccc|ccc} 
\hline
    \multicolumn{10}{c}{$n=30,m=5$}      \\
\hline
  $(\sigma_w^2+\sigma_b^2)/\rho^2$   &  \multicolumn{3}{c|}{0.2}   &  \multicolumn{3}{c|}{0.4}  
  &  \multicolumn{3}{c}{0.8}      \\
\hline
 $\sigma_w^2:\sigma_b^2$     &  
    1:3 & 1:1 & 3:1 &  1:3 & 1:1 & 3:1 & 1:3 & 1:1 & 3:1   \\
  \hline
     \multicolumn{10}{c}{$\alpha=0.05$}      \\
\hline
  GT     & 0.047 & 0.047  & 0.048  & 0.039  & 0.046  & 0.039  & 0.036 & 0.035  
          & 0.031  \\ 
  Z-score & 0.042 & 0.042  & 0.045 & 0.034  & 0.042 & 0.038 & 0.035 & 0.036 
          & 0.039  \\ 
  Z-Wald & 0.082 & 0.081  & 0.075 & 0.089  & 0.090 & 0.079 & 0.105 & 0.103 
          & 0.091  \\ 
\hline
     \multicolumn{10}{c}{$\alpha=0.01$}      \\
\hline
  GT     & 0.0087 & 0.0098  & 0.0097  & 0.0078  & 0.0095  & 0.0074  & 0.0073 & 0.0060  
          & 0.0048  \\ 
  Z-score & 0.0040 & 0.0051  & 0.0059 & 0.0030  & 0.0043 & 0.0042 & 0.0016 & 0.0019 
          & 0.0024  \\ 
  Z-Wald & 0.030 & 0.028  & 0.027 & 0.034  & 0.037 & 0.027 & 0.050 & 0.045 
          & 0.034  \\ 
\hline
    \multicolumn{10}{c}{$n=30,m=10$}      \\
\hline
  $(\sigma_w^2+\sigma_b^2)/\rho^2$   &  \multicolumn{3}{c|}{0.2}   &  \multicolumn{3}{c|}{0.4}  
  &  \multicolumn{3}{c}{0.8}      \\
\hline
 $\sigma_w^2:\sigma_b^2$     &  
    1:3 & 1:1 & 3:1 &  1:3 & 1:1 & 3:1 & 1:3 & 1:1 & 3:1   \\
  \hline
     \multicolumn{10}{c}{$\alpha=0.05$}      \\
\hline
  GT     & 0.045 & 0.045  & 0.047  & 0.048  & 0.044  & 0.044  & 0.037 & 0.036  
          & 0.038  \\ 
  Z-score & 0.040 & 0.041  & 0.044 & 0.042  & 0.039 & 0.041 & 0.035 & 0.035 
          & 0.041  \\ 
  Z-Wald & 0.081 & 0.073  & 0.071 & 0.097  & 0.090 & 0.079 & 0.106 & 0.103 
          & 0.092  \\ 
\hline
     \multicolumn{10}{c}{$\alpha=0.01$}      \\
\hline
  GT     & 0.0083 & 0.0095  & 0.0085  & 0.0096  & 0.0089  & 0.0083  & 0.0060 & 0.0059  
          & 0.0062  \\ 
  Z-score & 0.0037 & 0.0055  & 0.0046 & 0.0031  & 0.0036 & 0.0038 & 0.0016 & 0.0022 
          & 0.0029  \\ 
  Z-Wald & 0.029 & 0.027  & 0.023 & 0.042  & 0.035 & 0.026 & 0.052 & 0.044 
          & 0.036  \\ 
\hline
    \multicolumn{10}{c}{$n=30,m=20$}      \\
\hline
  $(\sigma_w^2+\sigma_b^2)/\rho^2$   &  \multicolumn{3}{c|}{0.2}   &  \multicolumn{3}{c|}{0.4}  
  &  \multicolumn{3}{c}{0.8}      \\
\hline
 $\sigma_w^2:\sigma_b^2$     &  
    1:3 & 1:1 & 3:1 &  1:3 & 1:1 & 3:1 & 1:3 & 1:1 & 3:1   \\
  \hline
     \multicolumn{10}{c}{$\alpha=0.05$}      \\
\hline
  GT     & 0.048 & 0.049  & 0.044  & 0.040  & 0.044  & 0.042  & 0.035 & 0.034  
          & 0.038  \\ 
  Z-score & 0.044 & 0.044  & 0.041 & 0.035  & 0.039 & 0.039 & 0.033 & 0.033 
          & 0.038  \\ 
  Z-Wald & 0.086 & 0.079  & 0.068 & 0.092  & 0.073 & 0.108 & 0.103 & 0.094 
          & 0.105  \\ 
\hline
     \multicolumn{10}{c}{$\alpha=0.01$}      \\
\hline
  GT     & 0.0110 & 0.0099  & 0.0074  & 0.0082  & 0.0084  & 0.0079  & 0.0065 & 0.0073  
          & 0.0055  \\ 
  Z-score & 0.0053 & 0.0042  & 0.0042 & 0.0031  & 0.0029 & 0.0033 & 0.0022 & 0.0027 
          & 0.0023  \\ 
  Z-Wald & 0.031 & 0.028  & 0.020 & 0.035  & 0.034 & 0.027 & 0.050 & 0.044 
          & 0.038  \\ 
\hline
\end{tabular}
\end{adjustbox}
\end{table}

\clearpage

\subsection{Power}

For the unbalanced study, we consider testing $H_0: \rho>3$.
For the balanced study, we set $\rho_0=\sqrt{6}$ and consider testing $H_0: \rho>3$. 
We use $10^4$ Monte Carlo simulations to estimate power under each configuration. 
Tables S4-S6 summarize the power.
Overall, we have observed very similar conclusions across simulation scenarios.
The proposed GT performs better than the Z-score test by a large margin. 
For the proposed GT,
(1) smaller between-subject variance $\sigma_b^2$
leads to larger power;
and
(2) larger $\mu$ (smaller $\sigma_w^2+\sigma_b^2$) also leads to larger power.
\begin{table}[h]
\caption{Power (\%) at significance level $\alpha$ estimated over $10^4$ simulations.
Data are simulated for $n=10$ subjects each with $m$ measures under $\sigma_w^2+\sigma_b^2+\mu^2=6$.
We are testing $H_0: \rho\geq 3$.
GT is the proposed test, and Z-score is the score based Z-test. }
\centering
\label{pwr1}
\begin{adjustbox}{max width=1\textwidth}
\begin{tabular}{c|ccc|ccc|ccc} 
\hline
    \multicolumn{10}{c}{$n=10,m=5$}      \\
\hline
  $(\sigma_w^2+\sigma_b^2)/\rho^2$   &  \multicolumn{3}{c|}{0.2}   &  \multicolumn{3}{c|}{0.4}  
  &  \multicolumn{3}{c}{0.8}      \\
\hline
 $\sigma_w^2:\sigma_b^2$     &  
    1:3 & 1:1 & 3:1 &  1:3 & 1:1 & 3:1 & 1:3 & 1:1 & 3:1   \\
  \hline
     \multicolumn{10}{c}{$\alpha=0.05$}      \\
\hline
  GT     & 42.7 & 53.2  & 68.7  & 25.6  & 31.1  & 44.3  & 14.7 & 18.9  
          & 26.0  \\ 
  Z-score & 33.2 & 43.7  & 61.7 & 17.5  & 23.2 & 37.3 & 9.7 & 14.0 
          & 23.8  \\ 
\hline
     \multicolumn{10}{c}{$\alpha=0.01$}      \\
\hline
  GT     & 15.0 & 21.5  & 33.9  & 7.1  & 9.5  & 16.4  & 3.3 & 5.1
          & 6.8  \\ 
  Z-score & 0.5 & 0.9  & 3.0 & 0  & 0.1 & 0.6 & 0 & 0 
          & 0.2  \\ 
\hline
    \multicolumn{10}{c}{$n=10,m=10$}      \\
\hline
  $(\sigma_w^2+\sigma_b^2)/\rho^2$   &  \multicolumn{3}{c|}{0.2}   &  \multicolumn{3}{c|}{0.4}  
  &  \multicolumn{3}{c}{0.8}      \\
\hline
 $\sigma_w^2:\sigma_b^2$     &  
    1:3 & 1:1 & 3:1 &  1:3 & 1:1 & 3:1 & 1:3 & 1:1 & 3:1   \\
  \hline
     \multicolumn{10}{c}{$\alpha=0.05$}      \\
\hline
  GT     & 43.5 & 57.2  & 79.0  & 25.9  & 34.9  & 54.2  & 15.1 & 21.4  
          & 36.1  \\ 
  Z-score & 33.9 & 47.7  & 71.8 & 18.0  & 25.6 & 44.9 & 9.6 & 15.2 
          & 29.1  \\ 
\hline
     \multicolumn{10}{c}{$\alpha=0.01$}      \\
\hline
  GT     & 15.4 & 24.2  & 44.7  & 7.6  & 11.4  & 22.2  & 3.6 & 5.7
          & 11.9  \\ 
  Z-score & 0.5 & 1.3  & 5.5 & 0.1  & 0.1 & 0.7 & 0 & 0 
          & 0.1  \\ 
\hline
    \multicolumn{10}{c}{$n=10,m=20$}      \\
\hline
  $(\sigma_w^2+\sigma_b^2)/\rho^2$   &  \multicolumn{3}{c|}{0.2}   &  \multicolumn{3}{c|}{0.4}  
  &  \multicolumn{3}{c}{0.8}      \\
\hline
 $\sigma_w^2:\sigma_b^2$     &  
    1:3 & 1:1 & 3:1 &  1:3 & 1:1 & 3:1 & 1:3 & 1:1 & 3:1   \\
  \hline
     \multicolumn{10}{c}{$\alpha=0.05$}      \\
\hline
  GT     & 44.7 & 59.4  & 83.8  & 26.5  & 36.9  & 60.5  & 15.9 & 23.5  
          & 44.0  \\ 
  Z-score & 34.7 & 49.5  & 76.8 & 18.3  & 27.3 & 50.8 & 10.4 & 15.8 
          & 33.8  \\ 
\hline
     \multicolumn{10}{c}{$\alpha=0.01$}      \\
\hline
  GT     & 16.6 & 26.0  & 51.3  & 7.6  & 12.4  & 26.8  & 3.7 & 6.3
          & 15.6  \\ 
  Z-score & 0.5 & 1.7  & 7.1 & 0  & 0.2 & 1.0 & 0 & 0 
          & 0.1  \\ 
\hline
\end{tabular}
\end{adjustbox}
\end{table}

\begin{table}[h]
\caption{Power (\%) at significance level $\alpha$ estimated over $10^4$ simulations.
Data are simulated for $n=20$ subjects each with $m$ measures under $\sigma_w^2+\sigma_b^2+\mu^2=6$.
We are testing $H_0: \rho\geq 3$.
GT is the proposed test, and Z-score is the score based Z-test. }
\centering
\label{pwr1}
\begin{adjustbox}{max width=1\textwidth}
\begin{tabular}{c|ccc|ccc|ccc} 
\hline
    \multicolumn{10}{c}{$n=20,m=5$}      \\
\hline
  $(\sigma_w^2+\sigma_b^2)/\rho^2$   &  \multicolumn{3}{c|}{0.2}   &  \multicolumn{3}{c|}{0.4}  
  &  \multicolumn{3}{c}{0.8}      \\
\hline
 $\sigma_w^2:\sigma_b^2$     &  
    1:3 & 1:1 & 3:1 &  1:3 & 1:1 & 3:1 & 1:3 & 1:1 & 3:1   \\
  \hline
     \multicolumn{10}{c}{$\alpha=0.05$}      \\
\hline
  GT     & 74.6 & 84.9  & 95.6  & 49.6  & 61.3  & 78.9  & 32.0 & 42.6  
          & 60.7  \\ 
  Z-score & 71.1 & 82.5  & 95.0 & 45.4  & 57.6 & 76.8 & 29.6 & 40.6 
          & 61.1  \\ 
\hline
     \multicolumn{10}{c}{$\alpha=0.01$}      \\
\hline
  GT     & 43.0 & 58.3  & 80.4  & 21.6  & 30.3  & 49.0  & 10.3 & 16.1
          & 28.4  \\ 
  Z-score & 24.3 & 38.7  & 65.4 & 7.9  & 14.1 & 30.2 & 2.5 & 4.6 
          & 13.1  \\ 
\hline
    \multicolumn{10}{c}{$n=20,m=10$}      \\
\hline
  $(\sigma_w^2+\sigma_b^2)/\rho^2$   &  \multicolumn{3}{c|}{0.2}   &  \multicolumn{3}{c|}{0.4}  
  &  \multicolumn{3}{c}{0.8}      \\
\hline
 $\sigma_w^2:\sigma_b^2$     &  
    1:3 & 1:1 & 3:1 &  1:3 & 1:1 & 3:1 & 1:3 & 1:1 & 3:1   \\
  \hline
     \multicolumn{10}{c}{$\alpha=0.05$}      \\
\hline
  GT     & 75.4 & 88.0  & 98.2  & 51.2  & 66.2  & 87.8  & 33.0 & 48.1  
          & 73.2  \\ 
  Z-score & 72.1 & 85.9  & 97.7 & 46.5  & 62.2 & 86.2 & 30.5 & 45.0 
          & 71.6  \\ 
\hline
     \multicolumn{10}{c}{$\alpha=0.01$}      \\
\hline
  GT     & 44.7 & 62.8  & 89.0  & 21.2  & 34.1  & 61.4  & 11.0 & 19.5
          & 41.0  \\ 
  Z-score & 25.9 & 43.4  & 78.5 & 7.6  & 16.2 & 41.7 & 2.4 & 5.5 
          & 18.9  \\ 
\hline
    \multicolumn{10}{c}{$n=20,m=20$}      \\
\hline
  $(\sigma_w^2+\sigma_b^2)/\rho^2$   &  \multicolumn{3}{c|}{0.2}   &  \multicolumn{3}{c|}{0.4}  
  &  \multicolumn{3}{c}{0.8}      \\
\hline
 $\sigma_w^2:\sigma_b^2$     &  
    1:3 & 1:1 & 3:1 &  1:3 & 1:1 & 3:1 & 1:3 & 1:1 & 3:1   \\
  \hline
     \multicolumn{10}{c}{$\alpha=0.05$}      \\
\hline
  GT     & 77.1 & 90.0  & 99.4  & 51.5  & 69.2  & 91.8  & 33.7 & 50.6  
          & 81.4  \\ 
  Z-score & 74.1 & 88.0  & 99.2 & 47.2  & 65.3 & 90.1 & 31.1 & 47.2 
          & 79.1  \\ 
\hline
     \multicolumn{10}{c}{$\alpha=0.01$}      \\
\hline
  GT     & 46.4 & 66.7  & 93.2  & 22.4  & 36.9  & 69.3  & 11.3 & 21.4
          & 51.4  \\ 
  Z-score & 26.9 & 46.9  & 84.7 & 8.4  & 17.3 & 49.3 & 2.4 & 5.9 
          & 25.2  \\ 
\hline
\end{tabular}
\end{adjustbox}
\end{table}

\begin{table}[h]
\caption{Power (\%) at significance level $\alpha$ estimated over $10^4$ simulations.
Data are simulated for $n=30$ subjects each with $m$ measures under $\sigma_w^2+\sigma_b^2+\mu^2=6$.
We are testing $H_0: \rho\geq 3$.
GT is the proposed test, and Z-score is the score based Z-test. }
\centering
\label{pwr1}
\begin{adjustbox}{max width=1\textwidth}
\begin{tabular}{c|ccc|ccc|ccc} 
\hline
    \multicolumn{10}{c}{$n=30,m=5$}      \\
\hline
  $(\sigma_w^2+\sigma_b^2)/\rho^2$   &  \multicolumn{3}{c|}{0.2}   &  \multicolumn{3}{c|}{0.4}  
  &  \multicolumn{3}{c}{0.8}      \\
\hline
 $\sigma_w^2:\sigma_b^2$     &  
    1:3 & 1:1 & 3:1 &  1:3 & 1:1 & 3:1 & 1:3 & 1:1 & 3:1   \\
  \hline
     \multicolumn{10}{c}{$\alpha=0.05$}      \\
\hline
  GT     & 90.0 & 96.4  & 99.5  & 68.1  & 80.5  & 93.5  & 47.0 & 61.8  
          & 80.2  \\ 
  Z-score & 88.5 & 95.9  & 99.4 & 65.6  & 78.8 & 93.1 & 45.8 & 61.0 
          & 80.2  \\ 
\hline
     \multicolumn{10}{c}{$\alpha=0.01$}      \\
\hline
  GT     & 67.4 & 83.3  & 96.0  & 37.0  & 52.7  & 75.3  & 19.3 & 30.9
          & 51.1  \\ 
  Z-score & 55.0 & 74.2  & 93.1 & 23.1  & 38.0 & 64.1 & 9.0 & 17.8 
          & 37.8  \\ 
\hline
    \multicolumn{10}{c}{$n=30,m=10$}      \\
\hline
  $(\sigma_w^2+\sigma_b^2)/\rho^2$   &  \multicolumn{3}{c|}{0.2}   &  \multicolumn{3}{c|}{0.4}  
  &  \multicolumn{3}{c}{0.8}      \\
\hline
 $\sigma_w^2:\sigma_b^2$     &  
    1:3 & 1:1 & 3:1 &  1:3 & 1:1 & 3:1 & 1:3 & 1:1 & 3:1   \\
  \hline
     \multicolumn{10}{c}{$\alpha=0.05$}      \\
\hline
  GT     & 90.7 & 97.4  & 99.95  & 68.8  & 84.2  & 97.2  & 49.0 & 67.2  
          & 91.1  \\ 
  Z-score & 89.5 & 97.1  & 99.95 & 66.2  & 82.5 & 96.8 & 47.5 & 65.9 
          & 90.7  \\ 
\hline
     \multicolumn{10}{c}{$\alpha=0.01$}      \\
\hline
  GT     & 68.7 & 86.7  & 98.8  & 39.1  & 57.9  & 85.7  & 20.6 & 35.8
          & 68.9  \\ 
  Z-score & 56.7 & 79.1  & 97.5 & 24.8  & 42.8 & 77.2 & 10.0 & 19.8 
          & 53.1  \\ 
\hline
    \multicolumn{10}{c}{$n=30,m=20$}      \\
\hline
  $(\sigma_w^2+\sigma_b^2)/\rho^2$   &  \multicolumn{3}{c|}{0.2}   &  \multicolumn{3}{c|}{0.4}  
  &  \multicolumn{3}{c}{0.8}      \\
\hline
 $\sigma_w^2:\sigma_b^2$     &  
    1:3 & 1:1 & 3:1 &  1:3 & 1:1 & 3:1 & 1:3 & 1:1 & 3:1   \\
  \hline
     \multicolumn{10}{c}{$\alpha=0.05$}      \\
\hline
  GT     & 91.8 & 97.9  & 99.98  & 70.3  & 86.2  & 98.6  & 49.5 & 71.1  
          & 95.0  \\ 
  Z-score & 90.6 & 97.6  & 99.96 & 67.7  & 84.6 & 98.3 & 47.5 & 69.3 
          & 94.5  \\ 
\hline
     \multicolumn{10}{c}{$\alpha=0.01$}      \\
\hline
  GT     & 71.3 & 88.3  & 99.5  & 39.1  & 60.0  & 91.2  & 21.2 & 40.4
          & 77.7  \\ 
  Z-score & 59.3 & 81.1  & 98.9 & 25.4  & 45.6 & 84.5 & 10.1 & 23.2 
          & 61.9  \\ 
\hline
\end{tabular}
\end{adjustbox}
\end{table}

\clearpage

\subsection{CI calculation}
Table S7-S9 summarize the coverage probability (CP)
and average width (AW) of computed 90\% CIs over $10^4$ simulations. 
Overall we can see that the CIs produced by the proposed method (denoted as GCI) 
have very good coverage probability close to the nominal 90\% level.
As expected, the Z-score test based approach has more conservative
performance, with CP generally much larger than the desired 90\% level and
CI much wider than the GCI. 
\begin{table}[h]
\caption{Coverage probability (CP) and average width (AW) of 90\% CI estimated over $10^4$ simulations.
Data are simulated from a balanced study with $n=20$ subjects each with $m=10$ measures and $\rho=3$.
GCI is the proposed test, Z-score is the score based Z-test. }
\centering
\label{err}
\begin{adjustbox}{max width=1\textwidth}
\begin{tabular}{c|ccc|ccc|ccc} 
\hline
    \multicolumn{10}{c}{$n=20,m=5$}      \\
\hline
  $(\sigma_w^2+\sigma_b^2)/\rho^2$   &  \multicolumn{3}{c|}{0.2}   &  \multicolumn{3}{c|}{0.4}  
  &  \multicolumn{3}{c}{0.8}      \\
\hline
 $\sigma_w^2:\sigma_b^2$     &  
    1:3 & 1:1 & 3:1 &  1:3 & 1:1 & 3:1 & 1:3 & 1:1 & 3:1   \\
  \hline
     \multicolumn{10}{c}{CP}      \\
\hline
  GCI     & 0.891 & 0.892  & 0.905  & 0.885 & 0.889  & 0.897  & 0.874 & 0.881  
          & 0.878  \\ 
  Z-score & 0.939 & 0.934  & 0.935 & 0.958  & 0.950 & 0.947 & 0.980 & 0.975 
          & 0.965  \\ 
\hline
     \multicolumn{10}{c}{AW}      \\
\hline
  GCI     & 1.259 & 1.080  & 0.891  & 1.689  & 1.427  & 1.170  & 2.209 & 1.804  
          & 1.437  \\ 
  Z-score & 1.324 & 1.109  & 0.888 & 1.873  & 1.523 & 1.200 & 2.373 & 1.828 
          & 1.393  \\ 
\hline
    \multicolumn{10}{c}{$n=20,m=10$}      \\
\hline
  $(\sigma_w^2+\sigma_b^2)/\rho^2$   &  \multicolumn{3}{c|}{0.2}   &  \multicolumn{3}{c|}{0.4}  
  &  \multicolumn{3}{c}{0.8}      \\
\hline
 $\sigma_w^2:\sigma_b^2$     &  
    1:3 & 1:1 & 3:1 &  1:3 & 1:1 & 3:1 & 1:3 & 1:1 & 3:1   \\
  \hline
     \multicolumn{10}{c}{CP}      \\
\hline
  GCI     & 0.895 & 0.892  & 0.893  & 0.888 & 0.889  & 0.895  & 0.883 & 0.882  
          & 0.883  \\ 
  Z-score & 0.941 & 0.933  & 0.921 & 0.958  & 0.953 & 0.944 & 0.981 & 0.976 
          & 0.964  \\ 
\hline
     \multicolumn{10}{c}{AW}      \\
\hline
  GCI     & 1.242 & 1.036  & 0.791  & 1.655  & 1.354  & 1.020  & 2.134 & 1.660  
          & 1.195  \\ 
  Z-score & 1.297 & 1.056  & 0.783 & 1.827  & 1.428 & 1.036 & 2.287 & 1.653 
          & 1.144  \\ 
\hline
    \multicolumn{10}{c}{$n=20,m=20$}      \\
\hline
  $(\sigma_w^2+\sigma_b^2)/\rho^2$   &  \multicolumn{3}{c|}{0.2}   &  \multicolumn{3}{c|}{0.4}  
  &  \multicolumn{3}{c}{0.8}      \\
\hline
 $\sigma_w^2:\sigma_b^2$     &  
    1:3 & 1:1 & 3:1 &  1:3 & 1:1 & 3:1 & 1:3 & 1:1 & 3:1   \\
  \hline
     \multicolumn{10}{c}{CP}      \\
\hline
  GCI     & 0.893 & 0.901  & 0.902  & 0.888 & 0.890  & 0.898  & 0.885 & 0.884  
          & 0.884  \\ 
  Z-score & 0.939 & 0.939  & 0.927 & 0.959  & 0.953 & 0.941 & 0.981 & 0.979 
          & 0.964  \\ 
\hline
     \multicolumn{10}{c}{AW}      \\
\hline
  GCI     & 1.229 & 1.006  & 0.741  & 1.635  & 1.309  & 0.946  & 2.115 & 1.591  
          & 1.064  \\ 
  Z-score & 1.282 & 1.020  & 0.731 & 1.803  & 1.382 & 0.952 & 2.250 & 1.575 
          & 1.008  \\ 
\hline
\end{tabular}
\label{CI2}
\end{adjustbox}
\end{table}

\begin{table}[h]
\caption{Coverage probability (CP) and average width (AW) of 90\% CI estimated over $10^4$ simulations.
Data are simulated from a balanced study with $n=20$ subjects each with $m=10$ measures and $\rho=3$.
GCI is the proposed test, Z-score is the score based Z-test. }
\centering
\label{err}
\begin{adjustbox}{max width=1\textwidth}
\begin{tabular}{c|ccc|ccc|ccc} 
\hline
    \multicolumn{10}{c}{$n=20,m=5$}      \\
\hline
  $(\sigma_w^2+\sigma_b^2)/\rho^2$   &  \multicolumn{3}{c|}{0.2}   &  \multicolumn{3}{c|}{0.4}  
  &  \multicolumn{3}{c}{0.8}      \\
\hline
 $\sigma_w^2:\sigma_b^2$     &  
    1:3 & 1:1 & 3:1 &  1:3 & 1:1 & 3:1 & 1:3 & 1:1 & 3:1   \\
  \hline
     \multicolumn{10}{c}{CP}      \\
\hline
  GCI     & 0.899 & 0.896  & 0.900  & 0.898 & 0.898  & 0.900  & 0.885 & 0.885  
          & 0.890  \\ 
  Z-score & 0.933 & 0.928  & 0.922 & 0.948  & 0.943 & 0.936 & 0.963 & 0.962 
          & 0.952  \\ 
\hline
     \multicolumn{10}{c}{AW}      \\
\hline
  GCI     & 0.852 & 0.734  & 0.600  & 1.125  & 0.958  & 0.779  & 1.391 & 1.139  
          & 0.911  \\ 
  Z-score & 0.887 & 0.756  & 0.608 & 1.212  & 1.013 & 0.805 & 1.496 & 1.184 
          & 0.927  \\ 
\hline
    \multicolumn{10}{c}{$n=20,m=10$}      \\
\hline
  $(\sigma_w^2+\sigma_b^2)/\rho^2$   &  \multicolumn{3}{c|}{0.2}   &  \multicolumn{3}{c|}{0.4}  
  &  \multicolumn{3}{c}{0.8}      \\
\hline
 $\sigma_w^2:\sigma_b^2$     &  
    1:3 & 1:1 & 3:1 &  1:3 & 1:1 & 3:1 & 1:3 & 1:1 & 3:1   \\
  \hline
     \multicolumn{10}{c}{CP}      \\
\hline
  GCI     & 0.899 & 0.898  & 0.902  & 0.900 & 0.896  & 0.895  & 0.890 & 0.892  
          & 0.896  \\ 
  Z-score & 0.933 & 0.929  & 0.923 & 0.951  & 0.939 & 0.929 & 0.963 & 0.963 
          & 0.953  \\ 
\hline
     \multicolumn{10}{c}{AW}      \\
\hline
  GCI     & 0.841 & 0.701  & 0.535  & 1.109  & 0.909  & 0.687  & 1.357 & 1.055  
          & 0.761  \\ 
  Z-score & 0.874 & 0.720  & 0.539 & 1.187  & 0.956 & 0.707 & 1.450 & 1.091 
          & 0.772  \\ 
\hline
    \multicolumn{10}{c}{$n=20,m=20$}      \\
\hline
  $(\sigma_w^2+\sigma_b^2)/\rho^2$   &  \multicolumn{3}{c|}{0.2}   &  \multicolumn{3}{c|}{0.4}  
  &  \multicolumn{3}{c}{0.8}      \\
\hline
 $\sigma_w^2:\sigma_b^2$     &  
    1:3 & 1:1 & 3:1 &  1:3 & 1:1 & 3:1 & 1:3 & 1:1 & 3:1   \\
  \hline
     \multicolumn{10}{c}{CP}      \\
\hline
  GCI     & 0.902 & 0.901  & 0.896  & 0.900 & 0.890  & 0.900  & 0.892 & 0.893  
          & 0.894  \\ 
  Z-score & 0.936 & 0.928  & 0.918 & 0.948  & 0.937 & 0.933 & 0.964 & 0.961 
          & 0.952  \\ 
\hline
     \multicolumn{10}{c}{AW}      \\
\hline
  GCI     & 0.835 & 0.684  & 0.501  & 1.093  & 0.881  & 0.636  & 1.336 & 1.008  
          & 0.677  \\ 
  Z-score & 0.866 & 0.699  & 0.504 & 1.172  & 0.925 & 0.652 & 1.425 & 1.044 
          & 0.686  \\ 
\hline
\end{tabular}
\label{CI2}
\end{adjustbox}
\end{table}

\begin{table}[h]
\caption{Coverage probability (CP) and average width (AW) of 90\% CI estimated over $10^4$ simulations.
Data are simulated from a balanced study with $n=30$ subjects each with $m=10$ measures and $\rho=3$.
GCI is the proposed test, Z-score is the score based Z-test. }
\centering
\label{err}
\begin{adjustbox}{max width=1\textwidth}
\begin{tabular}{c|ccc|ccc|ccc} 
\hline
    \multicolumn{10}{c}{$n=30,m=5$}      \\
\hline
  $(\sigma_w^2+\sigma_b^2)/\rho^2$   &  \multicolumn{3}{c|}{0.2}   &  \multicolumn{3}{c|}{0.4}  
  &  \multicolumn{3}{c}{0.8}      \\
\hline
 $\sigma_w^2:\sigma_b^2$     &  
    1:3 & 1:1 & 3:1 &  1:3 & 1:1 & 3:1 & 1:3 & 1:1 & 3:1   \\
  \hline
     \multicolumn{10}{c}{CP}      \\
\hline
  GCI     & 0.897 & 0.900  & 0.902  & 0.894 & 0.896  & 0.896  & 0.891 & 0.890  
          & 0.892  \\ 
  Z-score & 0.925 & 0.926  & 0.921 & 0.937  & 0.935 & 0.928 & 0.957 & 0.955 
          & 0.943  \\ 
\hline
     \multicolumn{10}{c}{AW}      \\
\hline
  GCI     & 0.690 & 0.592  & 0.483  & 0.908  & 0.772  & 0.627  & 1.101 & 0.900  
          & 0.720  \\ 
  Z-score & 0.715 & 0.608  & 0.490 & 0.966  & 0.810 & 0.649 & 1.174 & 0.941 
          & 0.740  \\ 
\hline
    \multicolumn{10}{c}{$n=30,m=10$}      \\
\hline
  $(\sigma_w^2+\sigma_b^2)/\rho^2$   &  \multicolumn{3}{c|}{0.2}   &  \multicolumn{3}{c|}{0.4}  
  &  \multicolumn{3}{c}{0.8}      \\
\hline
 $\sigma_w^2:\sigma_b^2$     &  
    1:3 & 1:1 & 3:1 &  1:3 & 1:1 & 3:1 & 1:3 & 1:1 & 3:1   \\
  \hline
     \multicolumn{10}{c}{CP}      \\
\hline
  GCI     & 0.899 & 0.901  & 0.903  & 0.899 & 0.892  & 0.895  & 0.893 & 0.890  
          & 0.895  \\ 
  Z-score & 0.925 & 0.924  & 0.920 & 0.940  & 0.928 & 0.925 & 0.954 & 0.950 
          & 0.943  \\ 
\hline
     \multicolumn{10}{c}{AW}      \\
\hline
  GCI     & 0.678 & 0.565  & 0.433  & 0.891  & 0.730  & 0.554  & 1.074 & 0.835  
          & 0.603  \\ 
  Z-score & 0.701 & 0.579  & 0.438 & 0.945  & 0.765 & 0.570 & 1.140 & 0.870 
          & 0.618 \\ 
\hline
    \multicolumn{10}{c}{$n=30,m=20$}      \\
\hline
  $(\sigma_w^2+\sigma_b^2)/\rho^2$   &  \multicolumn{3}{c|}{0.2}   &  \multicolumn{3}{c|}{0.4}  
  &  \multicolumn{3}{c}{0.8}      \\
\hline
 $\sigma_w^2:\sigma_b^2$     &  
    1:3 & 1:1 & 3:1 &  1:3 & 1:1 & 3:1 & 1:3 & 1:1 & 3:1   \\
  \hline
     \multicolumn{10}{c}{CP}      \\
\hline
  GCI     & 0.899 & 0.893  & 0.900  & 0.897 & 0.894  & 0.899  & 0.890 & 0.893  
          & 0.898  \\ 
  Z-score & 0.930 & 0.921  & 0.917 & 0.940  & 0.931 & 0.928 & 0.954 & 0.954 
          & 0.947  \\ 
\hline
     \multicolumn{10}{c}{AW}      \\
\hline
  GCI     & 0.673 & 0.550  & 0.404  & 0.883  & 0.710  & 0.512  & 1.058 & 0.798  
          & 0.537  \\ 
  Z-score & 0.695 & 0.563  & 0.408 & 0.936  & 0.742 & 0.526 & 1.125 & 0.831 
          & 0.550  \\ 
\hline
\end{tabular}
\label{CI2}
\end{adjustbox}
\end{table}

\clearpage

\section{R package}

The following are some sample codes to install and use the `RAMgt' R package.
\begin{verbatim}
 ## install/load the package
 library(RAMgt)
## Simulation example: how to get summary stats from data
A = rep(1:10, 5:14)
Y = 0.5+rnorm(length(A))+rnorm(length(unique(A)))[A]
ng = as.vector(table(A))
sse = sum(tapply(Y, A, var)*(ng-1), na.rm=TRUE)
mus = tapply(Y, A, mean)
GTrms(ng,mus,sse, alpha=0.05,rho0=2.5)
## Z-test
Zrms(ng,mus,sse, alpha=0.05,rho0=2.5)
\end{verbatim}

The oximetry comparison data can be analyzed to reproduce the analysis results reported in the main paper. 
\begin{verbatim}
## PO comparison study
ng = c(9, 10, 10, 10, 5, 10, 10, 10, 10, 10, 10, 10, 2, 10, 10, 10)
mus = c(-0.026,0.447,0.083,-0.103,-2.587,-0.61,0.04,-0.593, 0.963,0.643,
        -0.2,-1.337,-4.333,-2.807,0.563,-0.797)
sse = 221.037
set.seed(123)
GTrms(ng,mus,sse, alpha=0.1, rho0=3, Bmc=1e4)
## Z-test
Zrms(ng,mus,sse, alpha=0.1,rho0=3)
\end{verbatim}

\end{document}